# Dynamical Screening of Local Spin Moments at Metal−Molecule Interfaces

Sumanta Bhandary,* Emiliano Poli, Gilberto Teobaldi, and David D. O'Regan

**ABSTRACT:** Transition-metal phthalocyanine molecules have attracted considerable interest in the context of spintronics device development due to their amenability to diverse bonding regimes and their intrinsic magnetism. The latter is highly influenced by the quantum fluctuations that arise at the inevitable metal−molecule interface in a device architecture. In this study, we have systematically investigated the dynamical screening effects in phthalocyanine molecules hosting a series of transition-metal ions (Ti, V, Cr, Mn, Fe, Co, and Ni) in contact with the Cu(111) surface. Using comprehensive density functional theory plus Anderson's Impurity Model calculations, we show that the orbital-dependent hybridization and electron correlation together result in strong charge and spin fluctuations. While the instantaneous spin moments of the transition-metal ions are near atomic-like, we find that screening gives rise to considerable lowering or even quenching of these. Our results highlight the importance of quantum fluctuations in metal-contacted molecular devices, which may influence the results obtained from theoretical or experimental probes, depending on their possibly material-dependent characteristic sampling time-scales.

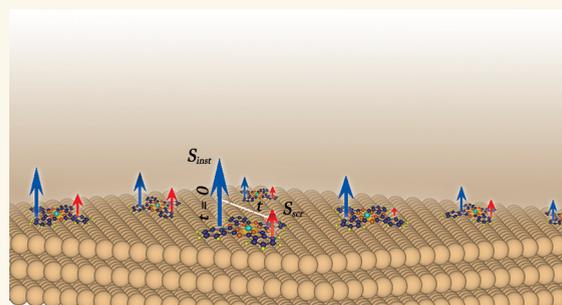

Advancement toward device development in the field of molecular spintronics[1−4] has its particular challenges. In spite of having an unmatched inherent size advantage and multiple exploitable intrinsic and extrinsic molecular features, such as switchable structural conformation,[5−8] charge state,[9,10] and magnetic properties,[11−15] robust device integration has remained a key bottleneck. In the context of spintronic applications, metal−organic molecules with transition-metal or rare-earth metal centers are an ostensibly ideal fit due to their intrinsic magnetism. Metallophthalocyanines or porphyrin molecules are of particular interest here, due to their ability to bond to varying degrees (e.g., in chemisorption or physisorption) with metallic and organic surfaces, in addition to their exploitable magnetism. While contact with a metallic surface is nearly inevitable in any device construction, it is also the main factor that influences the magnetic properties of the connected molecule, at times completely destroying them.[16,17] The degree of impact, however, relies largely on the type of surface. With storage device objectives, the device integration of molecules contacted with coinage metals such as Cu, Au, and Ag has been explored in depth, alongside those with ferromagnetic surfaces, such as Fe, Co, and Ni. Molecules often bond most strongly with ferromagnetic surfaces[18] and most weakly with organic layers.[19] In the intermediate case of the coinage metals, however, the couplings are strongly surface- and orientation-dependent.

Transition-metal ions, the carriers of magnetic properties in metallo-organic molecules, experience a different ligand field and a varying degree of itinerancy in contact with surfaces, hence laying a path for external manipulation. Indeed, modification of a surface−molecule interaction has been used to change the magnetic properties of adsorbed molecules, which can give rise, for example, to spin state,[18−20] magnetic coupling,[21,22] magnetic anisotropy,[23] or Kondo resonance.[24,25] Moreover, by coupling surface modification to external triggers, fundamental device functionalities have been demonstrated, such as spin-switches,[26−28] spin transistors,[29−32] and spin-valves.[33] A completely different approach for device development has emerged by means of organic embedding of metallo-organic molecules in two-dimensional (2D) systems.[34] Working in the opposite sense, organic molecules have been found to significantly affect the surface magnetization, in both the saturation magnetization and coercivity, of metallic substrates,[35,36] even turning conventional nonmagnets to weak ferromagnets,[37,38] leading to the field of emergent magnetism.

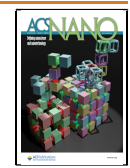

Table 1. Adsorption Energies and Preferred Adsorption Positions of the TMPc Molecules Analyzed on Cu(111) Surfaces

|  | Adsorption energies [eV] | | | | | | | |
| --- | --- | --- | --- | --- | --- | --- | --- | --- |
|  | TiPc | VPc | CrPc | MnPc | FePc | CoPc | NiPc | CuPc |
| Energy | −7.16 | −6.66 | −5.80 | −5.92 | −6.18 | −5.96 | −5.32 | −5.68 |
| Position | Top | Top | Top | Top | Top | Top | Top | Bridge |

As we move toward realistic device integration, the impact on the magnetic properties when molecular and inorganic subsystems come into proximity merits a detailed and cautious inspection. In the present work, we study how surface adsorption on Cu(111) impacts the magnetic properties of extensively studied transition-metal phthalocyanine molecules. In spite of being in the same group of the periodic table, the molecular bonding with Cu is often significantly different than that of Au and Ag.[39] Cu can lead to a suppression or enhancement of the magnetic properties of neighboring materials, and thus careful consideration is needed if its use is considered for device architectures, noting that the extrinsic control offered by Cu also depends strongly on relative surface orientations.[39]

The experimental technique chosen, it should be emphasized, can play a crucial role in the identification of magnetic properties such as the spin moment. When the coupling to the surface is large, such as in a chemisorbed interface, molecular spins tend to be influenced by dynamical screening effects stemming from the hybridization with the surface. This is particularly the case for metallic surfaces. A prime example is the appearance of a Kondo resonance peak.[24,40,41] Irrespective of the appearance of a Kondo peak, the mechanism via surface hybridization causes screening of different strengths that depends on surface, orientation, or molecular geometry. In bulk correlated-electron materials, such as Fe-based superconductors, varied dynamical screening has been seen.[42−44] However, the identification of this effect strongly relies on the time scale of the measurement techniques.[45] Fast probes such as X-ray absorption spectroscopy (XAS) and X-ray emission spectroscopy (XES) often show a discrepancy in the measured local moments as compared to those observed in inelastic neutron scattering (INS).[45] Moreover, recent work has shown that dynamic charge and spin fluctuations can affect the magnetic anisotropy.[46]

As the persistence at a high value of the (long-time) (following an experimental stimulus) magnetic moment, distinct from the instantaneous moment, is the key to spintronics applications, the dynamical screening effect in the local magnetic moment at longer time scales needs to be addressed in detail. This has been particularly lacking to date in the field of molecular magnetism. One of the advantages brought about by the coupling of density-functional theory (DFT) and many-body model Hamiltonian solvers via Green's functions, developed over the past several years, is that they allow for dynamical quantum fluctuations to be monitored and for the resulting screening to be explicitly analyzed in detail.

In this work, we utilize a physically realistic theoretical model combining semilocal approximate DFT with many-body theory in the paradigm of the multiorbital Anderson Impurity Model (AIM), to systematically analyze the dynamical screening effects in the local moments of transition-metal (TM) ions in transition-metal phthalocyanine (TMPc) on Cu(111) surface, as we scan across the full range of first-row transition-metal elements.

We show that a varying degree of hybridization is found for different TMs in TMPc, in contact with the Cu(111) surface specifically, and that this leads to the suppression of the high atomic-like moments to varying degrees. Crucially, concerning the hybridization and enabling the mechanistic understanding of this effect, these molecule−metal systems show strong orbital selectivity, making the process of screening highly orbital-dependent. The central features of the mechanism are that the molecular symmetry causes the $d_{x^2-y^2}$ orbital to strongly hybridize allowing a high degree of itinerancy; the $d_{xy}$ orbital remains highly nonbonding, promoting Hund's first rule physics, while the intermediate itinerancy in the out-of-plane orbitals shows strongly occupancy-dependent dynamical screening. As discussed in depth in the following sections, the variation in orbital itinerancy, combined with the 3d population and a strong Coulomb interaction, actually determines the retention, suppression, or quenching of the local moments. The emergence of this mechanistic understanding, through our discussion, will enable future steps toward design rules for maximizing local magnetic moments at metallo-organic interfaces, which is a critical necessity for hybrid spintronics applications.

## RESULTS AND DISCUSSION

**Molecular adsorption and structure.** The TMPc molecules were found to be weakly chemisorbed on the Cu(111) surface. All molecules, except for CuPc, favor the "top" site, meaning that the TM center of the molecules is positioned right on top of a Cu-surface atom. The CuPc molecule favors the bridge site, i.e., on the bond between two surface Cu atoms. It is to be noted that, for CuPc, the energy difference between the "top" and the "bridge" adsorption sites is very small, ∼4 meV, the latter site being favored. In a practical scenario, we expect to find a mixture of these two adsorption sites. For all the structures considered the adsorption energy ($E_{ads}$) was calculated as the binding energy between the TMPc molecule and Cu substrate.

$$E_{ads} = E_{tot} - E_{Cu} - E_{TMPc} \qquad (1)$$

Here, $E_{tot}$ is the energy of the optimized TMPc/Cu(111) system; $E_{Cu}$ refers to the optimized energy of the isolated Cu substrate, and $E_{TMPc}$ refers to the optimized energy of the isolated TMPc molecules. We use the convention where negative adsorption energies refer to a more stable composite system. The calculated adsorption energies for all of the systems considered for Anderson's Impurity model calculations are presented in Table 1.

The structural mismatch at the adsorption sites brings in local distortions both to the Cu-surface as well as to the molecules, while the latter is impacted more.

Due to a direct interaction between the out-of-plane orbitals of the TM and Cu, the transition-metal ion and the "top" Cu atoms are dragged closer to each other, yielding a convex structure of the molecular core (made of one TM, directly connected to four N atoms of the phthalocyanine molecule). A schematic representation of such a molecular distortion is presented in the inset of Figure 1. The degree of distortion from free-molecular planarity, however, depends on the covalent radius of the TM ion and the strength of the interaction with the surface. In Figure 1, we plot the metal–molecule distance, $d_{TM-Cu}$, and the distortion of the molecular core for all the TMPcs. To quantify the distortion, we have obtained the degree of nonplanarity ($\theta$) and the TM–N bond lengths, $d_{TM-N}$, as shown in Figure 1. We find that the larger the $\theta$ value, the larger the elongation of the TM–N bonds; this will have a strong impact on the intramolecular as well as molecule–surface interactions, as will be discussed later.

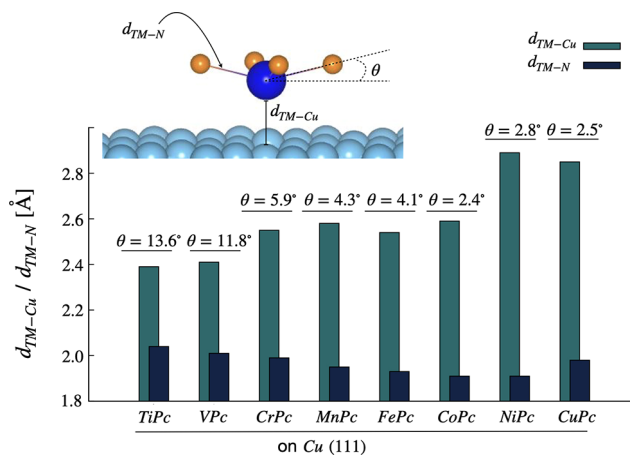

**Figure 1.** Structural details of the adsorbed transition-metal phthalocyanine molecules on the Cu(111) surface. $d_{TM-Cu}$ and $d_{TM-N}$ represent the closest distances between TM and surface Cu layer and intramolecular axial bonds between TM and N atoms, respectively. $\theta$ quantifies the degree of nonplanarity in the adsorbed molecules.

In the case of TiPc and VPc, the molecules are closest to the surface (∼2.4 Å), yielding strong transition-metal–Cu bonds, mediated through the out-of-plane 3d orbitals. Consequently, the degree of molecular distortion is high; i.e., $\theta = 13.6°$ and $\theta = 11.8°$, respectively, for TiPc and VPc. The Ti–N and V–N bond lengths are 2.04 and 2.01 Å, respectively.

The strong molecule–surface coupling can be attributed to the large covalent radii of Ti and V, which facilitate stronger chemical bond formation. As we travel rightward in the periodic table, the covalent radius decreases; hence, a weaker chemical bond can be expected. A jump in the TM–Cu bond distance (∼2.6 Å) is predicted for molecules containing Cr, Mn, Fe, and Co. As the chemical bonds are weakened, the molecules regain planarity with an ensuing reduction of the $\theta$ angle in the 5.9°–2.4° range. The molecule-to-surface distances for NiPc and CuPc are the highest, at ∼2.8 Å, with $\theta$ being 2.8° and 2.5°, respectively. The TM–N bond lengths of the molecular core strongly depend on the bond distortions; the larger the distortion, the longer the TM–N bond, which ranges from 2.04 Å for Ti to 1.91 Å for Ni.

**Hybridization.** The transition-metal ions in the molecules carry a propensity for magnetic moment formation that is strongly influenced by their hybridization with the phthalocyanine ring as well as with the surface underneath. To analyze these, we have studied the dynamic hybridization function (see eq S6 in the Supporting Information) of the 3d orbitals of the TM ions in the molecule plus surface environment, as detailed in the Supporting Information. In Figure 2, we plot the imaginary part of the hybridization function ($\Im\Delta$) (corresponding to the relevant 3d shell) for all TMPc molecules on Cu(111). It should be noted that $\Delta$ represents the hybridization of the 3d orbitals with both the molecular ring and the Cu-surface underneath. Within the 3d manifold, the $d_{x^2-y^2}$ orbital hybridizes strongly with the molecular ligand via axial overlap with 2p orbitals of the neighboring N atoms, a signature of molecules with square-planar symmetry.

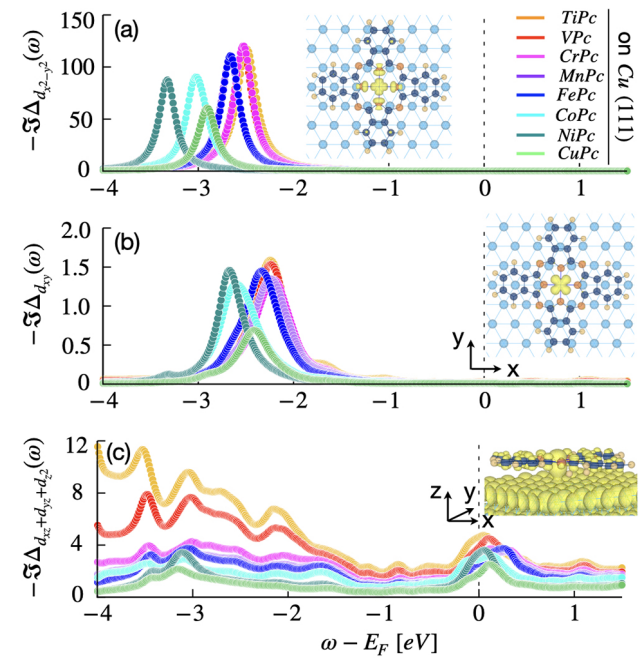

**Figure 2.** Orbital resolved dynamic hybridization function ($\Delta$) of different TMPc molecules on Cu(111). (a) Imaginary part of $\Delta_{d_{x^2-y^2}}$ for the TMPc molecules reflecting the strengths of the axial TM–N bonds. (b) Imaginary part of $\Delta_{d_{xy}}$, another in-plane 3d orbital. The intensities of $\Im(\Delta_{d_{xy}})$ are significantly smaller and lie far from the Fermi energy (−2.0 to −3.0 eV). (c) As per (b) but now for the sum of the out-of-plane orbitals ($d_{xz} + d_{yz} + d_{z^2}$). A broad $\Delta$ for the out-of-plane orbitals around the Fermi energy reflects the coupling of the TM orbitals to the Cu surface states.

In Figure 2a, we plot the strongest peak in $\Im\Delta_{d_{x^2-y^2}}$ for all of the TM ions. A sharp peak arises as a signature of molecular bond formation, in the energy range from −2.0 to −3.5 eV. We note that the intensity of a peak in $\Im\Delta$ is directly related to the coupling strength ($V_i$) (see eq S7 in the Supporting Information).

$V_i$ is stronger in the early TM ions, while it is weaker in the later part of the series, with the weakest being for Cu. This further illustrates the distortion in the TM–N$_4$ core discussed previously. On a relative scale, the couplings of the other orbitals are weak, a feature of their orbital symmetry and weak chemisorption on the metallic Cu-surface.

In Figure 2b,c, we plot $\Im\Delta_{d_{xy}}$ and $\Im\Delta_{d_{xz}+d_{yz}+d_{z^2}}$, respectively. The only significant molecular peak in $\Im\Delta_{d_{xy}}$ appears between

−1.5 and −3 eV. The weaker coupling and position far from the Fermi energy signifies a very weak hybridization with the molecular ligand. Indeed, the molecular symmetry only allows the weakest hybridization of the $d_{xy}$ orbitals, both with the molecular ligands as well as Cu surface, and hence the latter remain rather nonbonding in character. We note, however, that the $\Im\Delta_{d_{xy}}$ is slightly broader in adsorbed molecules compared to the same in the free TMPc molecules. This signifies a slight enhancement of the coupling, even in the $d_{xy}$ orbitals in the adsorbed molecules.

In contrast to the in-plane $d_{x^2-y^2}$ and $d_{xy}$ orbitals, the energy dependence of the out-of-plane orbitals ($\Im\Delta_{d_{xz}+d_{yz}+d_{z^2}}$) is broad. This illustrates a significant hybridization of the out-of-plane 3d orbitals with the metallic states of the Cu(111) surface. The degree of this itinerancy, however, is different for different adsorbed molecules. If scaled, the $d_{z^2}$ orbital shows more metallic behavior compared to the quasi-degenerate $d_{xz}$ and $d_{yz}$ orbitals, as one can see in Figure S1 in the Supporting Information. In Figure S1, it can also be noted that the $d_{z^2}$ hybridization gradually weakens as the atomic number increases. It thus follows that the set of molecules considered offers a realistic yet almost model-like tunability that enables the study of the role of itinerancy in dynamical screening.

Finally, we emphasize that the incorporation of electron correlation by many-body theory technique, for the systems considered in this paper, needs to go beyond discrete Hamiltonian approaches such as exact diagonalization. This is because of the relatively large and broad hybridization functions at the Fermi energy. Indeed, these effects are critical, as they can induce significant charge and spin fluctuations, which can result in the screening of molecular local magnetic moments or, potentially, Kondo resonance.

**Spin−Spin Correlation and Effective Local Moments.** The orbital occupations and the local magnetism are dictated by the electron correlation and the combined ligand field of the molecular ring and the Cu-surface underneath. Although these molecules exhibit significant local moments, both in crystal or in monolayer, they remain paramagnetic due to a far-too-weak magnetic exchange or single-ion anisotropy. However, the local moments can be harnessed for a variety of spintronic applications.[26−32] This has encouraged a large cohort of experimental and theoretical works that have explored the local moments in molecular crystals or adsorbed molecules on different surfaces, including Kondo resonance in specific molecule−surface hybrids. Away from the Kondo picture, magnetism in the molecule−surface remains an intricate puzzle,[47−50] and a systematic study illustrating dynamical screening impacting the molecular magnetic moments in different degrees has been missing in the field of molecular magnetism.

The electron itinerancy, which affects the local moment at the transition-metal ion in these metal−molecule hybrids, stems from (1) a strong intramolecular hybridization, i.e., the hybridization of the TM ion with the rest of the molecule, and (2) moderate direct (via direct orbital overlap) and indirect (via orbital overlap through molecular ligand states) hybridization with the Cu(111) surface. The latter varies strongly with the type of surface and the surface orientation. More importantly, this makes the 3d hybridization orbital-dependent, which, in turn, makes the screening process highly orbital-dependent, in strong contrast with the trend in bulk correlated materials.

To illustrate the effects of dynamical screening on local moments, we calculate the magnetic dipole moment ($M$) and hence the effective spin angular momentum ($S$), from the spin-susceptibility ($\chi$) within a multiorbital (five 3d orbitals) Anderson's Impurity Model (AIM), employing a continuous-time quantum Monte Carlo (CTQMC) solver (for calculation details, see Supporting Information).

$$\chi_{ij}(\tau) = g^2 \langle \hat{S}_i^z(\tau) \hat{S}_j^z(0) \rangle \tag{2}$$

$$M^2(\tau) = 3 \sum_{ij} \chi_{ij}(\tau) \tag{3}$$

Here, $\tau$ is the imaginary time, $i$ and $j$ are TM 3d orbitals, $\hat{S}_i^z$ is the z-component of the local spin operator, and $g$ is the spin gyromagnetic factor. We performed the CTQMC calculations presented in this paper at inverse temperature, $\beta = 1/k_BT = 40$ eV$^{-1}$, which corresponds to room temperature (290 K). $k_B$ is the Boltzmann constant. We obtained the effective spin angular momentum ($S$), from $M^2 = g^2 S(S+1)$, taking into account all the components of $\chi_{ij}(\tau)$. Specifically, we will be comparing and contrasting the instantaneous value $S_{inst} = S(\tau = 0)$ and the long-time, or screened, value $S_{scr} = S(\tau = \beta/2)$, inferred from $M(\tau)$.

The Coulomb interactions in such confined molecules are sizable, and consequently, the orbital-off diagonal components of the spin-susceptibility become important. We have considered a full Coulomb parametrization, which preserves the spin rotational invariance of the Coulomb tensor and where the magnetic moment can be given by

$$M^2 = g^2[\langle \hat{S}_x^2 \rangle + \langle \hat{S}_y^2 \rangle + \langle \hat{S}_z^2 \rangle] = 3g^2 \langle \hat{S}_z^2 \rangle \tag{4}$$

At $\tau = 0$, $\chi$ is proportional to the square of the bare local spin moment, i.e., the unscreened or instantaneous paramagnetic moment ($S_{inst}$), which is developed at a short time scale (~fs).[45] Over a longer time scale, quantum fluctuations tend to screen the local moment, often partially, at times fully quenching it. At $\tau = \beta/2$, $\chi$ corresponds to the effective moment at an asymptotically long time, incorporating the dynamical screening due to quantum fluctuations,[42,51] the "long time" or the screened spin moment ($S_{scr}$).

In Figure 3, we show the instantaneous or unscreened (purple bar) and screened (cyan bar) effective moments, $S_{inst}$ and $S_{scr}$, respectively, along with the spin moments obtained from the DFT calculation with static exchange and correlation (DFT+U) (green bar). We refer readers to the Supporting Information for the computational details.

It is interesting to note that all the TM ions in this molecule−surface environment show large atomic-like (indeed, slightly larger than atomic-like due to higher valence charge) instantaneous moments. At longer time scales, such large bare moments are screened significantly but to nontrivial and varying degrees. The screened moments show a reduction of and, in cases, even quenching of the local moments. $\Delta_s$ quantifies the difference between the bare and screened moments; the variation in $\Delta_s$ exhibits the dependence on the TM species embedded in the phthalocyanine molecules.

To illuminate the mechanism behind this surprisingly varied degree of moment screening, we identify: (1) the strong orbital dependence of the hybridization function and (2) the TM-dependent filling of the 3d shell, which is influenced significantly by the former. To show the ingredients of the

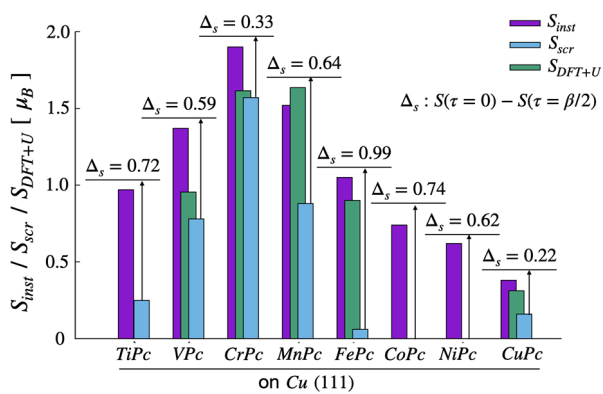

Figure 3. Dynamical screening of the effective moments in TMPc molecules on Cu(111). Unscreened (purple), screened (cyan), and DFT+U effective local moments (green) for TMPc molecules on Cu(111). $\Delta_s$ is the difference between the instantaneous and the screened spin moments.

mechanism up front, in Figure 4 we plot total spin susceptibility $\chi(\tau)$ (Figure 4a), orbital resolved occupations

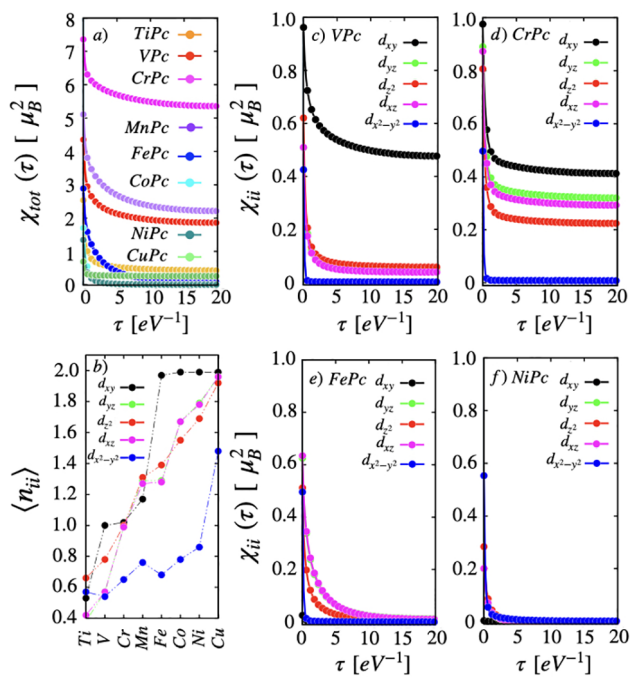

Figure 4. Orbital resolved spin susceptibility ($\chi_{ii}$) in the imaginary time $\tau$ for all TM ions in TMPc/Cu(111) systems. A different degree of screening is observed for different orbitals due to mainly two factors: (1) different strength of hybridization with the molecule and the surfaces and (2) impurity occupation.

(Figure 4b) for all the TMPc molecules, and orbital-diagonal elements of $\chi(\tau)$ (Figure 4c−f) of selected molecules, i.e., VPc, CrPc, FePc, and NiPc. We refer readers to the Supporting Information for all other molecules.

At this point, let us reiterate that, as a signature of the square-planar symmetry, the $d_{x^2-y^2}$ orbitals show the strongest intramolecular coupling. The spin in $d_{x^2-y^2}$ orbital experiences a metallic screening due to strong electron itinerancy that causes fluctuation of the local spin moment. This results in a sharp drop in $\chi(\tau)$ for the $d_{x^2-y^2}$ orbitals in all the TMPc molecules (Figure 4c−f). Hence the screened moment does not have any contribution from the $d_{x^2-y^2}$ orbital.

The magnetism is carried by the rest of the orbitals ($d_R$: $d_{xy}$, $d_{xz}$, $d_{yz}$, and $d_{z^2}$), which remain close in energy.[52] In this regard, CuPc is the only exception, where the $d_R$ subspace is fully filled and magnetism is carried by the $d_{x^2-y^2}$ orbital, as detailed later. It is to be noted that, despite the orbitals in $d_R$ being energetically closely spaced, $d_{xy}$ behaves significantly differently. The $d_{xy}$ orbital remains the weakest coupled orbital, hybridizing only weakly to the molecular ring or to the Cu surface states, as can be seen in Figure 2. Hence, the $d_{xy}$ orbital faces the weakest metallic screening and can contribute to the "long-time" moment, if partially occupied. The out-of-plane orbitals $d_{xz}$, $d_{yz}$, and $d_{z^2}$, due to their hybridization with the Cu-surface metallic states, exhibit stronger charge fluctuations in the out-of-plane orbital space, meaning a stronger suppression of $\chi(\beta/2)$, as seen in Figure 4. This is a clear distinction with the free molecules, as discussed in detail in the Supporting Information, where a very similar flattening of $\chi(\tau)$ can be noticed for all the orbitals in $d_R$ due to the absence of a surface underneath.

The local moments are partially screened in (Ti−Mn)Pc and CuPc, while they are strongly suppressed in FePc and fully quenched in (Co−Ni)Pc, in contrast with their conventional bulk ferromagnetic counterparts. A generic categorization of such varied degree of screening can be made on the basis of orbital filling in the $d_R$ subspace (and resultant quantum fluctuations), as follows.

*At Half-Filling.* The $d_R$ subshell is half-filled in CrPc, as one can see in Figure 4b, where $\langle n \rangle \approx 1$ for each orbital. In accordance, the $\chi(0)$ and corresponding $S_{inst}$ are the highest, dictated by the Hund's coupling. Weakened charge fluctuations and strong ferromagnetic spin fluctuations (as detailed in the following section) result in a significant remnant in $\chi(\tau)$ at $\tau = \beta/2$; therefore, CuPc shows the largest screened moment among the TMPc molecules. The effect of screening is the lowest in the CrPc molecule, where $\Delta_s$ amounts to only 17.4% of the instantaneous spin.

The orbital-diagonal terms of $\chi(\tau)$ offer further insights into the orbital-dependent nature of the screening. The weakly hybridized $d_{xy}$ orbital is almost half-filled with CrPc. In spite of a small metallic screening, the quantum fluctuations can cause screening of the spin in $d_{xy}$, as evidenced by the drop in $\chi(\tau)$. We note that there is a slight enhancement in the $d_{xy}$ hybridization for the molecules on the Cu-surface compared to that in the free molecules.

This reduces the atomic-like flattening of $\chi(\tau)$ of the $d_{xy}$ orbitals, unlike in the free molecules (as shown in Figure S7 in the Supporting Information), implying a slightly enhanced screening in the adsorbed molecules. Among the out-of-plane orbitals, the spin in the $d_{z^2}$ faces a comparatively larger screening, which can be attributed to a stronger hybridization with the surface (see Figure S1 in the Supporting Information). Each orbital in the $d_R$ subspace has a significant remnant in $\chi(\tau)$ at $\tau = \beta/2$, maximizing the long-term local spin moment in CrPc.

*Away from Half-Filling.* On both sides of the effective half-filling of $d_R$, the instantaneous moment shows an atomic-like decrease in Figure 3, evidenced by the average occupations, $\langle n_{ii} \rangle$, that indicate empty or doubly occupied orbitals reducing $S_{inst}$.

However, this deviation from effective "half-filling" has a far-reaching impact on the screened moment owing to enhanced charge fluctuations, as detailed in the following section. It thereby causes a decrease in both the instantaneous and long-time local moments.

In TiPc, VPc, and MnPc, $\chi(\tau)$ shows a significant drop at $\tau = \beta/2$. The corresponding $\Delta_s$ shows larger values as compared to that in CrPc, at 40−70% of the instantaneous spin moments. The half-filled $d_{xy}$ orbitals contribute significantly to the persisting local moment, while spins in the out-of-plane orbitals are strongly screened. Further away from the half-filling, in FePc, the $d_{xy}$ orbital is filled. The effective screened moment is strongly suppressed, almost quenching the local moment. Quenching of spin moments of FePc on Cu surfaces has been seen in recent experiments.[48,53] Upon lowering the temperature, the screening is significantly increased, further suppressing the Fe moment, in good agreement with the results of experiments performed at low temperatures. We refer the readers to Supporting Information for a detailed discussion on the temperature dependence of dynamical screening. In contrast, using DFT+U we obtained an $S \approx 1$ spin state on Fe.

*Quenching of Local Moments.* A special scenario appears in CoPc and NiPc. In spite of finite (although comparatively smaller) instantaneous moments, the $\chi(\tau)$ at $\beta/2$ there vanishes, and hence, also the screened moments vanish. In these cases, the $d_R$ subshell is much closer to being fully occupied, which, combined with the valence and charge fluctuations, quenches the local moments at a long time scale. While NiPc shows an $S \approx 0$ persisting (i.e., long-time) spin-state in both free (see Figure S6 in the Supporting Information) and adsorbed (on Cu) states, the $S = 1/2$ (persisting) local moment in CoPc, (almost) solely carried by the $d_{z^2}$ orbital, is experimentally found to be highly susceptible to screening on a variety of surfaces[54] including the Cu(111) surface.[55]

*Complete Filling.* Unlike other molecules, in CuPc the $d_R$ subspace is fully filled. An unpaired electron in the $d_{x^2-y^2}$ orbital predominantly contributes to the local moment in CuPc. The presence of the local spin in $d_{x^2-y^2}$ leads to a weakening of the Cu−N axial bonds, much like in the spin-crossover molecules.[34,52,56] The almost completely filled $d_R$ subshell and weakened hybridization in $d_{x^2-y^2}$ minimizes the screening effects, resulting in a sizable screened local moment in CuPc. Such a persisting local moment in CuPc, unlike in other TMPc molecules, is quite similar to the free molecule moment (see Supporting Information) and is hardly impacted by the presence of a surface underneath, as also observed in multiple experiments.[50,54,57]

We note that, in free TMPc molecules, in the absence of hybridization with the surface, $\chi(\tau)$ in the $d_R$ subspace shows an atomic-like flattening with a significant, large value for $\chi(\tau = \beta/2)$. Therefore, relatively large screened spin moments are found in all TMPc molecules, except for NiPc. We refer the reader to Supporting Information for a detailed discussion.

In spite of it being a static, ground-state theory, the DFT calculations (with interactions treated at the level of DFT+U) can sometimes predict local moments in good agreement with experiments and many-body calculations. The situation can hold true even with a high degree of correlation and localization, such as correlated insulators, where the charge fluctuations are highly suppressed, resulting in $\chi(0)$ and $\chi(\beta/2)$ being close in value. An example of such a scenario is CrPc, where the effects of screening are much reduced and the atomic-like high magnetic moment is governed by Hund's coupling. However, in the case of MnPc and FePc and also in TiPc, in an intermediate regime akin to that of "correlated metals", the disagreement with the ground-state theory is greater. Therefore, in multiple instances, the DFT+U approach fails to describe the experimentally observed local moments.[47−50]

**Charge & Spin Fluctuations.** To further illuminate the origin of the local moments in TMPc molecules on a Cu(111) surface, we obtained the generalized double occupations $\langle n_{i,\sigma} n_{j,\sigma'} \rangle$. In Figure 5, we plot the matrix heat maps of the $\langle n_{i,\sigma} n_{j,\sigma'} \rangle$ for parallel ($\sigma' = \sigma$) and antiparallel ($\sigma' = \bar{\sigma}$) spins. For $\sigma' = \sigma$, the diagonal elements represent the averaged orbital-resolved occupations, $\langle n_i \rangle$. We note that the $\langle n_{i,\sigma} n_{j,\sigma'} \rangle$ is symmetrized over the orbitals and spins. The half-filling, i.e., $\langle n_i \rangle = 0.5$, electron case signifies well-defined local moments. $\langle n_{i,\sigma} n_{j,\sigma} \rangle \gg \langle n_{i,\sigma} n_{j,\bar{\sigma}} \rangle$ for $i \neq j$ marks the development of large local moments, dictated by Hund's coupling. The $\langle n_{i,\sigma} n_{j,\sigma} \rangle \approx \langle n_{i,\sigma} n_{j,\bar{\sigma}} \rangle$ indicates enhanced charge fluctuation in the system. We recall that, for spintronics applications, large local moments and small screening are desirable.

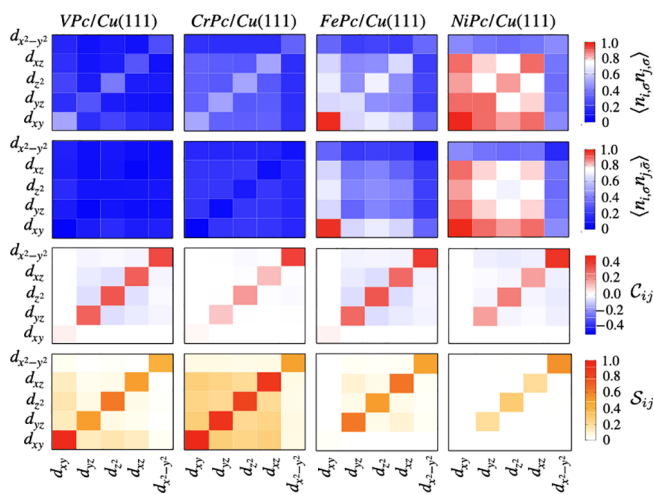

Figure 5. Heat-map representations of the (a) double occupation $\langle n_{i,\sigma} n_{j,\sigma} \rangle$ for like spins, (b) double occupation $\langle n_{i,\sigma} n_{j,\bar{\sigma}} \rangle$ for opposite spins, (c) charge- and (d) spin-fluctuation as defined in eqs 5 and 6.

A common feature, seen in all the adsorbed molecules except for CuPc, is that the $\langle n_{d_{x^2-y^2}} \rangle$ ranges between $\sim$0.3 and 0.4 electrons—due to the axial bond formation—which contributes to the instantaneous moment in a similar degree in all the molecules.

The $d_R$ subshell in VPc is far below half-filling. The $\langle n_{d_{xy}} \rangle$ is $\sim$0.5 electrons; $\langle n_{d_{z^2}} \rangle$ is $\sim$0.4 electrons, fairly close to the half-filling, while the $d_{xz}$ and $d_{yz}$ orbitals have relatively low occupations. Within $d_R$, the $\langle n_{i,\sigma} n_{j,\sigma} \rangle \gg \langle n_{i,\sigma} n_{j,\bar{\sigma}} \rangle$ between $d_{xy}$ and $d_{z^2}$, while $\langle n_{i,\sigma} n_{j,\sigma} \rangle \approx \langle n_{i,\sigma} n_{j,\bar{\sigma}} \rangle$ for the orbital off-diagonal terms involving $d_{xz}$ and $d_{yz}$ orbitals. These indicate a good localization in the $d_{xy}$ and $d_{z^2}$ orbitals with a strong Hund's coupling. The fluctuations are higher in the $d_{xz}$ and $d_{yz}$ orbitals.

The local moment is hence dominated by the $d_{xy}$ and $d_{z^2}$ orbitals, while $d_{xy}$ and $d_{yz}$ have lower contributions.

The situation of half-filling of the $d_R$ subspace occurs in the case of CrPc. Each orbital in $d_R$ has 0.5 electrons; at the same time, $\langle n_{i,\sigma} n_{j,\sigma} \rangle \gg \langle n_{i,\sigma} n_{j,\bar{\sigma}} \rangle$ for $i \neq j$ in the $d_R$ subspace, meaning a strong Hund's coupling prevails with significantly low charge fluctuations. These result in the highest instantaneous local moment in the CrPc molecule, $S_{inst} \approx 1.9$, a significant fraction of which is retained in the screened moment, $S_{scr} \approx 1.6$.

In MnPc, the $\langle n_{i,\sigma} n_{j,\sigma} \rangle$ for all orbitals in the $d_R$ subspace is ∼0.6 electrons, slightly over half-filling. The instantaneous spin moment hence is reduced to ∼1.5. In contrast to the CrPc molecule, the orbital off-diagonal terms for ($\sigma' = \sigma$) and ($\sigma' = \bar{\sigma}$) gain values of similar strengths. This signifies stronger charge fluctuations, which weaken the screened moment in MnPc. This further emphasizes the importance of the many-body description, and the discrepancy with DFT+U is much greater. It is important to note that, until MnPc in the TM series, a significant part of the local moment is carried by the half-filled (or partially filled) $d_{xy}$ orbital, which as mentioned before, is a nonbonding orbital. The metallic screening is strongly suppressed in this orbital; thus, it can contribute significantly to the screened moments.

Starting from FePc to NiPc, the $\langle n_{d_{xy}} \rangle = 1.0$, suppressing the local spin completely in that orbital. The out-of-plane orbitals in the $d_R$ subspace are slightly over half-filling in FePc, leading to $S_{inst} \approx 1$. Due to this increased occupation, the off-diagonal terms in $d_R$ subspace is relatively high for both ($\sigma' = \sigma$) and ($\sigma' = \bar{\sigma}$), a sign of weakened Hund's coupling. The charge fluctuations (both intra- and interorbital) are high, which causes the suppression of the screened local moments. The average orbital occupations in the out-of-plane orbitals are significantly increased in CoPc and NiPc, moving further away from half-filling, weakening the instantaneous moments to ∼0.7 and 0.6, respectively. The charge fluctuations are also strongly enhanced. It is to be noted that, in (Fe−Ni)Pc, the local moments stem from the out-of-plane orbitals because $d_{xy}$ is doubly occupied. Due to the hybridization with the metallic Cu states and simultaneously with higher orbital-filling, the screened local moments are suppressed to the highest degree.

This situation of local moment suppression can be further understood from the explicitly determined charge and spin correlations

$$C_{ij} = \langle \hat{n}_i \hat{n}_j \rangle - \langle \hat{n}_i \rangle \langle \hat{n}_j \rangle \quad (5)$$

$$\mathcal{S}_{ij} = \langle \hat{\sigma}_i \hat{\sigma}_j \rangle - \langle \hat{\sigma}_i \rangle \langle \hat{\sigma}_j \rangle \quad (6)$$

where the operators are $\hat{n}_i = \hat{n}_{i\uparrow} + \hat{n}_{i\downarrow}$ and $\hat{\sigma}_i = \hat{n}_{i\uparrow} - \hat{n}_{i\downarrow}$. It is to be noted that $\langle \hat{\sigma}_i \rangle = 0$ in the paramagnetic case.

Both the intraorbital ($C_{ii}$) and the interorbital ($C_{ij}$) are increased as the orbital occupation (absolute value) deviates from the half-filled or the fully filled scenario. Therefore, we see the weakest charge fluctuations in CrPc and NiPc, where $d_R$ is half-filled and (almost) filled, respectively. Such a picture can also be viewed in the orbital-resolved $C_{ii}$, namely, that the $d_{xy}$ orbital (with the least hybridization) shows the least fluctuation when half-filled (in VPc and CrPc) and fully occupied (in Fe-NiPc), while $d_{x^2-y^2}$ has almost similar values over the TM series, dominated mostly by the strong in-plane hybridization. The interorbital fluctuations involving $d_{x^2-y^2}$ are also insignificant, in accordance with the strong ligand-field separation from the $d_R$. Importantly, among the out-of-plane orbitals, both the intra- and interorbital fluctuations are strong (except for Cr and Ni, for the reason stated above), which is a signature of enhanced metallicity, as previously reported.[58]

The orbital-resolved spin fluctuations are presented in Figure 5 (lowest panel). The intraorbital terms ($\mathcal{S}_{ii}$) represent (are proportional to) the spin moments. In the $d_{x^2-y^2}$ orbital, a similar value over the TM series (except for CuPc) indicates an unchanged instantaneous moment. Higher charge fluctuations suppress the intraorbital correlator $\mathcal{S}_{ii}$, hence reducing the local moment. This can be vividly seen in CrPc, where weaker charge fluctuations lead to strong local moments in $d_R$. At the same time, high intraorbital correlators represent the stabilization of high-spin moments. In CrPc and MnPc, the interorbital (off-diagonal) correlation is the strongest, showing high local moments, among the TM series. As the orbital filling increases toward the right (Fe-NiPc), both the intra- and interorbital correlators are strongly suppressed, signifying strong screening of the local moments.

## CONCLUSIONS

In summary, we have analyzed in detail how metal−molecule contact, which is an inevitable part of molecular device design, can strongly impact molecular magnetism. Our study takes in the full range of first-row transition-metal ions in metal phthalocyanines, physisorbed or weakly chemisorbed on the Cu(111) surface. The TMPc molecules in contact with the Cu(111) surface show varied degrees of magnetic screening, significantly reducing and even quenching local magnetic moments. The origin can be traced back to electron correlation, and we have detailed a scheme for its rationalization in terms of the orbital-selective hybridization of TM 3d orbitals with the molecular ring as well as with metallic Cu states. In this picture, the effective ligand field controls the orbital filling and hence the formation of the local moments. A large instantaneous moment arises at the half-filling of the $d_R$ subshell, and, more crucially, quantum charge fluctuations are strongly suppressed at half-filling. Hence, large screened moments are seen in cases such as CrPc. In many other systems, a deviation from half-filling results in stronger quantum fluctuations, which are further enhanced by metallic hybridization with the Cu-surface and, ultimately, in strongly reduced or even quenched screened spin magnetic moments. Interestingly, we find that, for certain systems close to half-filling, the density self-consistent DFT+U prediction for the magnetic moment is in reasonably good agreement with the screened moment from the many-body solution of ostensibly the same impurity Hamiltonian. For other systems, DFT+U performs poorly in this regard, and we have explained why this discrepancy arises.

This study has further highlighted a possibly more important discrepancy between the predicted spin magnetic moment of the transition-metal ion sampled instantaneously and that at longer time scales over which screening by the metal substrate can take place. This distinction, which can amount to the qualitative observation versus nonobservation (complete quenching) of magnetic moments in some systems, may go some way to explain the differences in magnetic moment values observed in different experimental probes.[42−44] The strength of this difference will, of course, depend on the characteristic time-scales of fluctuating moments in molecule substrate hybrids and the intrinsic time scales of the specific experimental probe,[45] but we would expect some probes (e.g., XAS, XES) to sample moments over shorter time scales than others (e.g., inelastic neutron scattering). Overall, this study not only addresses the puzzle of vanishing moments in some recent studies[47−50] but also importantly provides a general guide, based on orbital filling and orientation considerations, for the rational design of nanodevices based on molecular magnetism on metallic surfaces.

## METHODS

We have performed calculations using combined first-principles plus many-body theory approaches for a realistic description of electronic structure and magnetism in metal−molecule hybrids. The first-principles calculations are based on density functional theory (DFT) employing a full potential plane wave-based program package, VASP.[59] We used the Perdew−Burke−Ernzerhof (PBE) generalized gradient approximation of exchange-correlation potential and included an empirical form of dispersion correction given by Grimme.[60] The simulation cell consists of an 8 × 8 lateral supercell of 3 Cu (111) layers, on which TMPc molecules are deposited. The size of the supercell is chosen in such a way that TM centers are at least 20 Å apart from each other, minimizing the interaction. In the vertical direction, we have considered a vacuum greater than 14 Å. The atomic positions are relaxed keeping the lowest Cu(111) layer fixed to the in-plane experimental lattice constant as of bulk Cu. The structures are relaxed until the Hellman Feynman forces are minimized below 0.01 eV/Å. We have used a 4 × 4 × 1 Monkhorst Pack k-mesh for the Brillouin zone integration. We note that, while the atomic-relaxations are spin-polarized calculations, we used nonspin-polarized calculations to extract parameters for the many-body calculations.

For the many-body calculations, we have used a continuous-time hybridization-expansion quantum Monte Carlo (QMC) solver as implemented in the w2dynamics program package.[61] To account for the strong electron correlation at the TM centers, we have considered rotationally invariant Coulomb interactions parametrized via the Slater radial integrals[62,63] $F^0$, $F^2$, and $F^4$, such that $U = F^0$ and $J = \frac{1}{14}(F^2 + F^4)$, with the ratio $F^4/F^2 = 0.625$, yielding a spherically symmetric tensor.[41,64] In all of our calculations, we have used $U = 4.0$ eV and $J = 1.0$ eV.

We refer the reader to the Supporting Information for a further detailed discussion on how to extract ab initio parameters through the dynamical hybridization function and on the many-body simulations.

## ASSOCIATED CONTENT

### Ⓢ Supporting Information

The Supporting Information is available free of charge at https://pubs.acs.org/doi/10.1021/acsnano.3c00247.

> Computational details of first-principles and many-body simulations; determination of the dynamical hybridization function from DFT calculation; comparison of hybridization functions of the out-of-plane 3d orbitals in different TMPc molecules on Cu(111); structural changes and charge transfer due to molecular adsorption; local moments and screening in adsorbed TMPc molecules on Cu(111), the temperature dependence of the dynamical screening; local moments and screening in isolated TMPc molecules (PDF)

## AUTHOR INFORMATION

### Corresponding Author

Sumanta Bhandary − *School of Physics and CRANN Institute, Trinity College Dublin, Dublin 2, Ireland;* orcid.org/0000-0001-7487-1720; Email: sumanta.bhandary@tcd.ie

### Authors

Emiliano Poli − *Scientific Computing Department, STFC UKRI, Rutherford Appleton Laboratory, Didcot OX11 0QX, United Kingdom*

Gilberto Teobaldi − *Scientific Computing Department, STFC UKRI, Rutherford Appleton Laboratory, Didcot OX11 0QX, United Kingdom; School of Chemistry, University of Southampton, Highfield SO17 1BJ Southampton, United Kingdom;* orcid.org/0000-0001-6068-6786

David D. O'Regan − *School of Physics and CRANN Institute, Trinity College Dublin, Dublin 2, Ireland;* orcid.org/0000-0002-7802-0322

Complete contact information is available at:
https://pubs.acs.org/10.1021/acsnano.3c00247

### Author Contributions

All authors contributed to the project design, results discussion, and reporting. S.B. and E.P. carried out the numerical simulations. S.B. created the figures. S.B. and D.O'R. interpreted the many-body calculations.

### Notes

The authors declare no competing financial interest.

## ACKNOWLEDGMENTS

We would like to thank Oscar Cespedes, Karsten Held, Angelo Valli, Jan Tomczak, and Alexander Kowalski for very insightful discussions. We acknowledge financial support from the Science Foundation Ireland [19/EPSRC/3605], the Engineering and Physical Sciences Research Council [EP/S031081/1 and EP/S030263/1]. DFT geometry optimizations have been carried on the ARCHER2 (accessed via membership of the UKCP consortium, EP/P022189/2 and EP/X035891/1) and STFC SCARF High Performance Computing facilities. The authors wish to acknowledge the Irish Centre for High-End Computing (ICHEC) for the provision of computational facilities and support.